\documentclass[showpacs,preprintnumbers,amsmath,amssymb]{revtex4}
\usepackage{graphicx}
\usepackage{dcolumn}
\makeatletter
\newcommand{\rb}[1]{\raisebox{1.5ex}[0pt]{#1}}
\input{epsf}
\begin{document}
\title{Novel Families of Fractional Quantum Hall States : A Berry Phase Approach}
\author{B. Basu}
 \email{banasri@isical.ac.in}

\author{P. Bandyopadhyay}
 \email{pratul@isical.ac.in}
\affiliation{Physics and Applied Mathematics Unit\\
 Indian Statistical Institute\\
 Kolkata-700108 }
\begin{abstract}
We have studied here the newly observed families of fractional
quantum Hall states in the framework of Berry phase. It has been
shown that this approach embraces in a unified way the whole
spectrum of quantum Hall states with their various characteristic
features. The newly observed states can be well accommodated
within the primary sequence of FQH states and need not be
considered as {\it second generation} FQH states as suggested by
some authors.
\end{abstract}
\maketitle
\section{introduction}
The recent experiments \cite{pan} on high mobility samples at very
low temperatures predicted the existence of fractional quantum
Hall (FQH) states at some unusual filling factors such as $\nu$
=4/11, 5/13, 5/17, 6/17, 4/13 and 7/11. For the states $\nu$=4/11
and 5/13 a deep minimum in $\rho_{xx}$ and also a respectable
plateau in $\rho_{xy}$ was observed. But for $\nu$=6/17, 4/13,
5/17 and 7/11 states there was no plateau in $\rho_{xy}$ and also
minimum in $\rho_{xx}$ was not pronounced. It has been observed
that the new state at $\nu$=4/11 is a fully polarized FQH state. A
weak state was also observed in $\nu$=3/8 and at $\nu$=3/10. These
sequences of fractions do not fit into the standard series
\cite{jain} of integral quantum Hall effects (IQHE) of composite
fermions (CF) at $\nu=\frac{p}{2mp \pm 1}$, but the states
$\nu=4/11~ {\rm or}~4/13$ appear in the hierarchy of quasiparticle
condensates \cite{hal,halp}. However, others such as $\nu=3/8~
{\rm or}~3/10$ do not belong to this hierarchy and the origin of
their incompressibility seems to be puzzling. It has been
proposed \cite{pan} that these states may be regarded as FQHE of
composite fermions attesting to residual interactions between
these composite particles.
 Finite
size diagonalization of small clusters of electrons by Mandal and
Jain \cite{man} and Chang et.al.\cite{chang} suggest that the
state at $\nu$=4/11 is partially polarized which differs from the
experimental findings of Pan et.al.\cite{pan}.

A significant analysis has been done by Quinn {\t et.
al.}\cite{quin1, quin2} in which they interpret these states as a
novel family of FQH states involving pairing correlations among
the quasiparticles (QP). The correlations depend upon the
behaviour of the QP-QP pseudopotential $V_{QP}(L')$ where the
interaction energy of a pair depends on the angular momentum.
These pairs are proposed  to have Laughlin correlations with one
another and to form condensed states at  a sequence which include
all new fractions found in experiments.

Besides, it is also proposed \cite{smith} that the new composite
particles consist each of a composite fermion of the first
generation and a vortex like excitation which is based on the
framework of Hamiltonian theory of Murthy and Shankar\cite{shan}.
Very recently, Lopez and Fradkin \cite{lopez} have proposed that
the new states may be viewed as the hierarchical Jain states such
that these are the quasiparticles and quasiholes of the primary
states of the Jain sequence. Indeed, they have constructed the FQH
states observed by Pan. et. al. as the fully polarized
hierarchical descendants of the Jain sereis.
All these approaches suggest that these new states effectively
represent the {\bf second generation} of FQH states as these are
expressed as FQH states of FQH states.

 In this note we shall show that all these new states may be
 treated as the FQH states in the primary sequence when we analyze
 the sereis from the viewpoint of Berry phase. Our analysis
 suggests the existence of FQH states which include apart from the
 states observed by Pan et. al. as well as all the predicted
 states by Quinn et. al. and also by Lopez and Fradkin.
Besides, in this formalism these states are found to be fully
polarized and their particle-hole conjugate states are found to be
unpolarized. This is consistent with the observation in
experiments for the state with $\nu=\frac{4}{11}$ and its
particle-hole conjugate state $\frac{7}{11}$.

In some earlier papers \cite{pb1, pb2} we have analyzed the
sequence of quantum Hall states from the viewpoint of chiral
anomaly and Berry phase. To this end, we have taken into account
quantum Hall states on the two dimensional surface of a $3$D
sphere with a magnetic monopole of strength $\mu$ at the centre.
In this spherical geometry we can analyze quantum Hall states in
terms of spinor wave functions and take advantage of the analysis
in terms of chiral anomaly which is associated with the Berry
phase. In this geometry the angular momentum relation is given by

\begin{equation}
{\bf J} ~ = ~ {\bf r} \times {\bf p} - \mu {\bf \hat{r}} ,
~~~~~~~~ \mu ~=~ 0 , ~\pm ~1 / 2 , ~\pm ~1 , ~\pm ~ 3 /
2,~.........
\end{equation}

From the description of spherical harmonics $Y^{m, \mu}_{\ell}$
with $\ell=1/2,~|m|=|\mu|=1/2$, we can construct a two-component
spinor $\theta = \left(
\begin{array}{c}
                          \displaystyle{u}\\
                          \displaystyle{v} \end{array} \right)$~~where
\begin{equation}
\begin{array}{lcl}
u ~=~ Y^{1/2 , 1/2}_{1/2} &=& \displaystyle{sin ~\frac{\theta}{2}
\exp
\left[ i (\phi - \chi) / 2 \right]}\\ \\
v ~=~ Y^{- 1/2 , 1/2}_{1/2} &=& \displaystyle{cos
~\frac{\theta}{2} \exp \left[- i (\phi + \chi) / 2 \right]}
\end{array}
\end{equation}
Here $\mu$ corresponds to the eigenvalue of the operator $i
\frac{\partial}{\partial \chi}$.

Then the $N$-particle wave function for the quantum Hall fluid
state at $\nu=\frac{1}{m}$ can be written as \cite{hal,pb1}
\begin{equation}
{\psi^{(m)}}_{N} ~=~ \prod_{i < j} {(u_i v_j - u_j v_i)}^{m}
\end{equation}
$m$ being an odd integer. Here $u_i(v_j)$ corresponds to the
$i$-th ($j$-th) position of the particle in the system.

It is noted that ${\psi^{(m)}}_{N}$ is totally antisymmetric for
odd $m$ and symmetric for even $m$. Following Haldane \cite{hal}
we can identify $m$ as $m=J_i+J_j$ for the N-particle system where
$J_i$ is the angular momentum of the $i$-th particle. It is
evident from eqn. (1) that with ${\bf r} \times {\bf p}=0$ and
$\mu=\frac{1}{2}$ we have
 $m=1$ which corresponds to  the complete
filling of the lowest Landau level.
 From the Dirac quantization condition $e
\mu =\frac{1}{2}$, we note that this state corresponds to $e=1$
describing the IQH state with $\nu=1$.

The next higher angular momentum state can be achieved either by
taking ${\bf r} \times {\bf p}=1$ and $\mid\mu \mid=\frac{1}{2}$
(which implies the higher Landau level) or by taking ${\bf r}
\times {\bf p}=0$ and $\mid{\mu_{eff}} \mid=\frac{3}{2}$ implying
the ground state for the Landau level. However, with
$\mid{\mu_{eff}} \mid=\frac{3}{2}$, we find the filling fraction
$\nu=\frac{1}{3}$ which follows from the condition $e \mu
=\frac{1}{2}$ for $\mu=\frac{3}{2}$. Generalizing this we can have
$\nu=\frac{1}{5}$ with $\mid{\mu_{eff}} \mid=\frac{5}{2}$.

It may be remarked that as $\mu$ here corresponds to the monopole
strength, we can relate this with the Berry phase. Indeed
$\mu=\frac{1}{2}$ corresponds to one flux quantum and the flux
through the sphere when there is a monopole of strength $\mu$ at
the centre, is $2\mu$ . The Berry phase of a fermion of charge $q$
is given by $e^{i\phi_B}$ with $\phi_B=2\pi q$ (number of flux
quanta enclosed by the loop traversed by the particle). It may be
mentioned that for a quantum Hall particle the charge is given by
$-\nu e$ when $\nu$ is the filling factor.

If $\mu$ is an integer, we can have a relation of the form
\begin{equation}
{\bf J} ~=~ {\bf r} \times {\bf p} - \mu {\bf \hat{r}} ~=~ {\bf
r}^{~\prime} \times {\bf p}^{~\prime}
\end{equation}
which indicates that the Berry phase associated with $\mu$ may be
unitarily removed to the dynamical phase. Evidently, the average
magnetic field may be considered to be vanishing in these states.
 The attachment of
$2m$ vortices ($m$ an integer) to an electron effectively leads to
the removal of Berry phase to the dynamical phase. So, FQH states
with  $2 \mu_{eff} = 2m + 1$  can be viewed as if one vortex line
is attached to the  electron. Now we note that for a higher Landau
level we can consider the Dirac quantization condition $e
\mu_{eff} = \frac{1}{2} n$, with $n$ being a vortex of strength $2
\ell + 1$. This can generate FQH states having the filling factor
of the form $\frac{n}{2 \mu_{eff}}$ where both $n$ and $2
\mu_{eff}$ are odd integers. Indeed, we can write the filling
factor as \cite{pb1,pb2}

\begin{equation}
\nu ~=~\frac{n}{2 \mu_{eff}} ~=~ \frac{1}{\frac{2 \mu_{eff}~ \mp
1}{n} \pm \frac{1}{n}} ~=~ \frac{n}{2mn \pm 1} \label{nu1}
\end{equation}
 where $2 \mu_{eff} \mp 1$ is an even integer
given by $2mn$.

In this scheme, the FQH states with $\nu$ having the form
\begin{equation}
\nu=\frac{n^{~\prime}}{2mn^{~\prime} \pm 1} \label{e2}
\end{equation}

($n^\prime$ an even integer) can be generated through
particle-hole conjugate states
\begin{equation}
\nu ~=~ 1 - \frac{n}{2mn \pm 1} ~=~ \frac{n (2m - 1) \pm 1}{2mn
\pm 1}=\frac{n^{\prime}}{2mn^{\prime} \pm 1} \label{e3}
\end{equation}

Now, from eqn.(5),
\begin{equation}
\frac{n}{2 \mu_{eff}} ~=~ \frac{n}{2m^{\prime}\pm 1} ~=~
\frac{n}{2mn \pm 1}
\end{equation}
we note that the integer $m^{\prime}$ has been taken to be the
product $mn$. However, not all $m'$ can be written in the product
form $mn$. So we can write in a generalized way,
$$m^{~\prime}=mn \pm \tilde{m},~~~~  \tilde{m}~~{\rm  being~an~integer}$$
Indeed replacing $n$ by $q$ where $q$ is any integer we can
express the generalized relation as
\begin{equation}
\frac{q}{2 \mu_{eff}} ~=~ \frac{q}{2m^{\prime}\pm 1} ~=~
\frac{n}{2mq \pm (2\tilde{m}\pm 1)}
\end{equation}
where we have taken $m^{~\prime}=mq\pm \tilde{m}$ , with
$\tilde{m}$ being an integer. The relation (\ref{nu1}) can be
expressed as
\begin{equation}\label{nu2}
\nu~=~\frac{1}{2m \pm \displaystyle{\left(\frac{2\tilde{m}\pm
1}{q}\right)}}~=~\frac{1}{2m \pm p/q}
\end{equation}
where $p\geq 1$ an odd integer. It is noted that for $p=1$
implying $\tilde{m}=0$ corresponds to the Jain sequence.

The even denominator filling factors are obtained when $\mu_{eff}$
is an integer. As we have pointed out that for integer $\mu$, the
Berry phase can be removed to the dynamical phase, these states
can only be observed when they appear as pair states \cite{pb3,
bdp}. This corresponds to the non-Abelian Berry phase and
represents non-Abelian quantum Hall fluid. In this case, the
relation (\ref{nu2}) will take the form
\begin{equation}\label{nu3}
\nu~=~\frac{1}{2m \pm
\displaystyle{\frac{2\tilde{m}}{q}}}~=~\frac{1}{2m \pm p/q}
\end{equation}
where $p~(q)$ is an even(odd) integer.

Now to have a physical interpretation of the states given by
(\ref{nu2}) and (\ref{nu3}) we note that these correspond to the
attachment of $p$ vortices (flux quanta) in a cluster of $q$
electrons in the lowest Landau level.

For $p>1$ and $q$ odd we can take \\
(i)  $q-2$ electrons of which each one is attached with a vortex
is coupled with a residual boson composed of two electrons \\
(ii) $q-1$ electrons, each of them being attached with a vortex is
coupled with the residual fermion.\\

In a similar way for $q$ even, we may view the relation
(\ref{nu2}) such that $q-1$ electrons, each having one vortex
attached with it is coupled with the residual fermion. It is noted
that when an electron is attached with a magnetic flux, its
statistics changes and it is transformed into a boson. These
bosons condense to form a cluster which is coupled with the
residual fermion or boson composed of two fermions. Indeed the
residual boson or fermion will undergo a "statistical" interaction
tied to a geometric Berry phase effect that winds the phase of the
particle as it encircles the vortices. This suggests that for odd
$q$, $p~$ takes the values $q-2$ and $q-1$ and for even $q$, $p~$
is given by $q-1$. Also we observe that the attachment of vortices
to electrons in a cluster will make the fluid an incompressible
one. Indeed as two vortices cannot be brought very close to each
other, there will be a hard core repulsion in the system which
accounts for the incompressibility of the quantum Hall fluid.

\newpage
Now we consider some specific cases:\\

{\underline{{\bf A :} $q$ odd, $p>1$ and $p=q-1$ or $q-2$}\\

\begin{tabular}{|c|c|c|c|c|}\hline
$q$&$p$&general form of $\nu$ &values of $m$&{$\nu$}\\  \hline & &
&1&{\bf 3/8}, 3/4 \\ \cline{4-5} \rb{3}&\rb{2}&\rb{$\frac{1}{2m
\pm
2/3}$}&2&3/14, {\bf 3/10} \\ \hline & & &1&{\bf 5/13}, {\underline{5/7}} \\
\cline{4-5}
 &\rb{3}&\rb{$\frac{1}{2m \pm 3/5}$}&2&5/23, {\bf 5/17} \\ \cline{2-5}
& & &1&5/14, 5/6\\ \cline{4-5}
\raisebox{4.0ex}[0pt]{5}&\rb{4}&\rb{$\frac{1}{2m \pm
4/5}$}&2&5/24, 5/16 \\ \hline
 & & &1&7/19, 7/9\\ \cline{4-5}
 &\rb{5}&\rb{$\frac{1}{2m \pm 5/7}$}&2&7/33, 7/23 \\ \cline{2-5}
\rb{7}& & &1&7/20, 7/8\\ \cline{4-5}
 &\rb{6}&\rb{$\frac{1}{2m \pm 6/7}$}&2&7/34, 7/22\\ \hline
\end{tabular}

\vspace{1cm}

{\underline{{\bf B :} $q$ even, $p>1$ and odd}\\

\begin{tabular}{|c|c|c|c|c|}\hline
$q$&$p$&general form of $\nu$ &values of $m$&{$\nu$}\\ \hline & &
&1&4/11, \underline{4/5} \\ \cline{4-5}
\rb{4}&\rb{3}&\rb{$\frac{1}{2m \pm 3/4}$}&2&4/19, 4/13, \\
\hline & & &1&{\bf 6/17}, 6/7 \\ \cline{4-5}
\rb{6}&\rb{5}&\rb{$\frac{1}{2m \pm 5/6}$}&2&6/29, {\bf 6/19} \\
\hline
\end{tabular}
\vspace{1cm}

Values with bold faces in the above Tables have been reported in
ref. [1]. The values $\nu=4/5$ and $\nu=5/7$ have been observed
by Du. et. al.\cite{dust} which appear as weak depressions in the
longitudinal resistivity. In a similar way, we can carry on with
other $q$ and $p$-values.

In this context we may add that the states we have obtained
include all the states predicted by Quinn et al.\cite{quin1,
quin2}. Indeed, their classification scheme suggests the relations
\begin{equation}
\nu^{-1}~=~2\tilde{p}~+~1~\pm \left( 2+\tilde{q}/2 \right)^{-1}
\end{equation}
where $\tilde{p}$ and $\tilde{q}$ are integers.

As mentioned earlier, the even denominator states are expected to
appear as paired states. Indeed, in this case $\mu_{eff}$ is an
integer and hence can be removed to the dynamical phase. These
states can only be observed in paired states. This suggests that
the newly observed states $\nu=3/8$ and $\nu=3/10$ should appear
in paired states which has also been suggested by some other
authors \cite{sca, quin1, quin2}. From our analysis, it appears
that these states correspond to non-Abelian Berry phase and
represent non-Abelian quantum Hall fluid.

In a recent paper \cite{bsp} we have analyzed the polarization of
quantum Hall states in the framework of the hierarchical analysis
in terms of the Berry phase. There it is observed that the states
in the lowest Landau level corresponding to the filling factors
$\nu=1$ and $\nu= \frac{1}{2m+1}$ with $m$ an integer correspond
to the fully polarized states. Indeed, in such a system even
number of vortices are gauged away and the attachment of one
vortex (magnetic flux) to an electron leads to the fully polarized
state. However, in the higher Landau level we have the filling
factor given by
$$\nu ~=~\frac{n}{2 \mu_{eff}} ~=~ \frac{1}{\frac{2 \mu_{eff}~ \mp
1}{n} \pm \frac{1}{n}} ~= \frac{1}{2m \pm \frac{1}{n}}=~
\frac{n}{2mn \pm 1}$$ where $n$ is an odd integer corresponding to
a vortex of of strength $2l+1$. As this effectively corresponds to
the attachment of $\frac{1}{n}$ flux unit attached to an electron,
the state will be partially polarized. Again the particle-hole
conjugate states given by
$$\nu ~=~ 1 - \frac{n}{2mn \pm 1},$$
n being an odd integer will correspond to unpolarized states. This
analysis suggests that the newly observed states as reported in
\cite{pan} which appear in the lowest Landau level will correspond
to fully polarized states. Indeed, the states corresponding to
$\nu=4/11$ has been observed to be fully polarized and the
particle-hole conjugate state $\nu=7/11$ is found to be
unpolarized \cite{pan}.

Finally, we may mention here that the hierarchical interpretation
of the Haldane-Halperin scheme was questioned \cite{quin3} because
of the specific form of the QP-QP interaction. Besides, as
mentioned earlier, the newly observed states $\nu=3/8$ and
$\nu=3/10$ do not belong to this hierarchy. On the other hand, the
Jain classification scheme effectively reveals a fundamental
connection between IQHE and FQHE as the FQH states are considered
to be IQH states of composite fermions . However, to interpret the
newly observed states this fundamental concept has to be abandoned
and we have to take into account the residual CF-CF interaction. A
specific form of this interaction \cite{quin1, quin2} based on the
pseudopotentials such that the pair interaction energy depends on
the relative angular momentum nicely explains the new states. But
some other states observed by Du. et.al \cite{dust} which appear
as weak depression in the longitudinal resistivity cannot be
accommodated in this scheme. Our present analysis suggests that
these {\it second generation} FQH states are not truly second
generation states, rather these appear in the primary sequence of
the FQH states. This classification scheme can explain all the
states observed to date \cite{pan, dust}. This also predicts the
polarization of the states $\nu=4/11$ and $\nu=7/11$ consistent
with experiment.

\end{document}